\documentclass[aoas,MSNbibl,nameyear,dvips]{arximspdf}
\usepackage{dcolumn}
\usepackage{graphicx}

\doi{10.1214/11-AOAS469}
\volume{5}
\issue{3}
\pubyear{2011}
\firstpage{2052}
\lastpage{2077}

\makeatletter

\let\sv@tabnotetext\tabnotetext
  \let\sv@tabnotemark@fmt\tabnotemark@fmt
   \long\def\legend#1{{\let\tabnote@indent\leavevmode\sv@tabnotetext[]{}{#1}}}

\def\@bmisc[#1]{%
  \get@battribute{unstr}%
  \common@pub@types%
  \let\bauthor\bbl@bauthor%
  \let\bhowpublished\@firstofone%
  \def\borganization##1{{\bauthor@style ##1}}%
}

\newcolumntype{d}[1]{D{,}{\ }{#1}}
\makeatother

\begin{document}
\begin{frontmatter}

\title{Block-based Bayesian epistasis association mapping
with application to WTCCC type 1 diabetes data\thanksref{TITL}}

\runtitle{Block-based Bayesian epistasis association mapping}

\thankstext{TITL}{This study makes use of data generated by the
Wellcome Trust Case--Control Consortium. A full list of the investigators
who contributed to the generation of the data is available from
\protect\href{http://www.wtccc.org.uk}{www.wtccc.org.uk}. Funding for the project was provided by the Wellcome
Trust under Award 076113. All the chromosomal positions are in NCBI
build 35 coordinates.}

\begin{aug}
\author[A]{\fnms{Yu} \snm{Zhang}\corref{}\ead[label=e1]{yuzhang@stat.psu.edu}\thanksref{aut1}},
\author[B]{\fnms{Jing} \snm{Zhang}\ead[label=e2]{jing.zhang.jz349@yale.edu}\thanksref{aut2}}
and
\author[C]{\fnms{Jun S.} \snm{Liu}\ead[label=e3]{jliu@stat.harvard.edu}\thanksref{aut2}}

\thankstext{aut1}{Supported by NIH Grant R01-HG004718.}
\thankstext{aut2}{Supported in part by the NIH Grant R01-HG02518-02 and the NSF Grant DMS-07-06989.}

\runauthor{Y. Zhang, J. Zhang and J. S. Liu}

\affiliation{Pennsylvania State University, Yale
University and Harvard University}

\address[A]{Y. Zhang\\
Department of Statistics\\
Pennsylvania State University\\
422A Thomas\\
University Park, Pennsylvania 16802\\
USA\\
\printead{e1}} 
\address[B]{J. Zhang\\
Department of Statistics\\
Yale University\\
24 Hillhouse Ave.\\
New Haven, Connecticut 06511\\
USA\\
\printead{e2}}
\address[C]{J. Liu\\
Department of Statistics\\
Harvard University\\
Science Center\\
1 Oxford St.\\
Cambridge, Massachusetts 02138\\
USA\\
\printead{e3}}
\end{aug}

\received{\smonth{4} \syear{2010}}
\revised{\smonth{12} \syear{2010}}

\begin{abstract}
Interactions among multiple genes across the genome may contribute to
the risks of many complex human diseases. Whole-genome single nucleotide
polymorphisms (SNPs) data collected for many thousands of SNP markers
from thousands of individuals under the case--control design promise to
shed light on our understanding of such interactions. However, nearby
SNPs are highly correlated due to linkage disequilibrium (LD) and the
number of possible interactions is too large for exhaustive evaluation.
We propose a novel Bayesian method for simultaneously partitioning SNPs
into LD-blocks and selecting SNPs within blocks that are associated with
the disease, either individually or interactively with other SNPs. When
applied to homogeneous population data, the method gives posterior
probabilities for LD-block boundaries, which not only result in accurate
block partitions of SNPs, but also provide measures of partition
uncertainty. When applied to case--control data for association mapping,
the method implicitly filters out SNP associations created merely by LD
with disease loci within the same blocks. Simulation study showed that
this approach is more powerful in detecting multi-locus associations
than other methods we tested, including one of ours. When applied to the
WTCCC type 1 diabetes data, the method identified many previously known
T1D associated genes, including \textit{PTPN22}, \textit{CTLA4}, MHC,
and \textit{IL2RA}. The method also revealed some interesting two-way
associations that are undetected by single SNP methods. Most of the
significant associations are located within the MHC region. Our analysis
showed that the MHC SNPs form long-distance joint associations over
several known recombination hotspots. By controlling the haplotypes of
the MHC class II region, we identified additional associations in both
MHC class I (\textit{HLA-A}, \textit{HLA-B}) and class III regions
(\textit{BAT1}). We also observed significant interactions between genes
\textit{PRSS16}, \textit{ZNF184} in the extended MHC region and the MHC
class II genes. The proposed method can be broadly applied to the
classification problem with correlated discrete covariates.
\end{abstract}


\end{frontmatter}

\section{Introduction}\label{sec1}

A recent genome-wide association (GWA) study of\break 14,000 cases of seven
human genetic diseases and 3,000 shared controls by the Welcome Trust
Case Control Consortium [WTCCC (\citeyear{m07})] represents a thorough validation
of the GWA approach. By testing hundreds of thousands of single
nucleotide polymorphisms (SNPs) in the human population (Affymetrix 500k
SNP), the study has identified many SNPs associated with seven complex
diseases [WTCCC (\citeyear{m07})]. Each SNP consists of two \textit{alleles}
taking values 0 or 1, and there are three possible combinations of the
two alleles: $(0,0)$, $(0,1)$, $(1,1)$, disregarding the order. Each
combination is called a~\textit{genotype}, representing wild homozygote,
heterozygote, and mutant homozygote, respectively. In a case--control
study, a SNP is said to be associated with the disease if the genotype
(or allele) distribution at the SNP is different between cases and
controls. In addition to testing individual SNPs, it has also been
anticipated that epistatic interactions among SNPs, defined as multiple
SNPs jointly associated with the disease, may be responsible for
significantly elevating the risks of some human complex diseases. Due to
computational and methodological limitations, however, efforts on
detecting disease-related epistatic interactions among SNPs in the WTCCC
study have been limited.

In the past few years, many approaches have been developed for
case--control studies to detect epistasis associations. Most methods
cannot be applied to GWA studies due to their computational limitations
except for some recently developed methods, such as the stepwise
logistic regression method [Marchini, Donnelly and Cardon (\citeyear{MarDonCar05})] and
the Bayesian epistasis association mapping (BEAM) algorithm [Zhang and
Liu (\citeyear{ZhaLiu07})]. It has been demonstrated that BEAM is capable of detecting
high-order interactions in GWA studies and is more powerful than other
existing methods [Zhang and Liu (\citeyear{ZhaLiu07})]. A limitation of BEAM, however,
is its model assumption that a Markov chain can capture the dependence
structure of the SNPs in the data. It is well known that linkage
disequilibrium (LD) between adjacent SNPs exhibits block-wise structure
in the human genome [The International HapMap Consortium (\citeyear{autokey22}), Reich et al. (\citeyear{Reietal01})]. SNPs
within blocks are highly correlated and the correlation is broken down
by recombination events at block boundaries. A simple Markov model
cannot capture this important block structure when analyzing dense
SNPs.\vadjust{\eject}

Previous studies have shown that using \textit{haplotypes}, defined as
allele combinations over multiple nearby SNPs inherited from one of the
parents, cannot only reduce the high computation cost in GWA studies,
but also improve the detection power of association mapping
[Kuno et al. (\citeyear{Kunetal04}); Z\"{o}llner and Pritchard (\citeyear{ZolPri05});
Johnson et al. (\citeyear{Johetal01}); Zhang et al. (\citeyear{Zhaetal02N1}); de Bakker et al. (\citeyear{deBetal05})]. In particular,
Nielsen et al. (\citeyear{Nieetal04}) showed that when moderate to high
levels of LD exist, haplotype tests tend to be substantially more
powerful. Kuno et al. (\citeyear{Kunetal04}) demonstrated in a real disease study that
single-SNP tests were not significant even at SNP loci close to the
mutation site (APRT*J), whereas the haplotype block data yielded
sufficient statistical significance.

Similar to haplotypes, we define \textit{diplotypes} as genotype
combinations over multiple nearby SNPs. Diplotypes are directly observed
in GWA studies, whereas haplotypes have to be inferred using
computationally expensive algorithms. Throughout the paper, we discuss
our method and results on diplotype associations with the disease,
although the method is directly applicable to haplotype data. Our focus
on diplotype association is mainly due to the computational concern,
where inferring unobserved haplotypes will be extremely time consuming.
It is also possible that testing diplotype associations could be more
powerful than testing haplotype associations, depending on the
underlying disease model. On the other hand, if haplotype associations
are of the interests, users can first infer haplotypes using an
available haplotype inference algorithm, and then input the inferred
haplotypes into our method. The degrees of freedom of our model will
automatically accommodate the different inputs.

In this paper we extend the BEAM model to address the block structures
in the human genome. We refer to SNP block structures as LD-blocks. By
partitioning SNPs into LD-blocks, a na\"{\i}ve extension of BEAM is to
treat each LD-block as a genetic marker, with diplotypes in the block
being treated as alleles. This approach, however, may not be optimal for
association mapping. First, criteria utilized in existing
block-partitioning algorithms do not directly aim at optimizing the
power of association mapping. Second, many regions in the human genome
demonstrate vague structural patterns, of which a measure of uncertainty
in block structures should be provided. Simulation studies have shown
that LD-block structures can be affected by marker density
[Wang et al. (\citeyear{Wanetal02}); Pillips et al. (\citeyear{Phietal03});
Wall and Pritchard (\citeyear{WalPri03})], population structure [Wang et al. (\citeyear{Wanetal02});
Stumpf and Goldstein (\citeyear{StuGol03}); Zhang et al. (\citeyear{Zhaetal03}); Anderson
and Slatkin (\citeyear{AndSla04})], and gene conversion [Przeworski and Wall (\citeyear{PrzWal01})].

We propose a Bayesian model to simultaneously infer LD-blocks and select
SNPs within blocks for disease association mapping. The model partitions
the genome into discrete blocks, within which the diversity of
diplotypes is limited. Block structures are iteratively updated
such
that disease associations are detected and summarized from a variety of\vadjust{\eject}
likely block partitions. This approach takes into account the
uncertainty in block structures and optimizes the detection power by
searching for the best partitions around disease-associated SNPs. Our
method detects combinations of SNPs within and between blocks for
marginal and epistatic associations to the disease status. Using
LD-blocks, our method also automatically filters out artificial
associations created merely by LD with nearby authentically associated
SNPs. We show that the new method, BEAM2, is more powerful than the
original BEAM (renamed BEAM1 henceforth).

By applying BEAM2 to the type 1 diabetes (T1D) data from WTCCC (\citeyear{m07}),
we obtained all the previously identified single SNP associations in the
WTCCC T1D data. We further observed some interesting two-way joint
associations not detectable by single-SNP methods. The strongest T1D
associations occur in the well-known Major Histocompatibility Complex
(MHC) region, in which we observed long-distance joint association
patterns over several millions of base pairs (Mb). Since this pattern
may be partially caused by the extended LD from the MHC class II region,
we further controlled the structure of MHC class II using a logistic
regression model, and tested additional effects of SNPs over the
extended MHC region as well as the MHC class I and class III regions. We
observed strong associations in the MHC class I and class III regions,
and found significant interaction associations between genes
\textit{PRSS16}, \textit{ZNF184} in the extended MHC region and the MHC
class II genes.

The article is organized as follows. In Section \ref{sec2} we first introduce a
LD-block model for the genotype distribution among multiple SNPs. We
model the genotype distribution within a block of SNPs by multinomial
distributions and assume that the joint distribution of all SNPs is the
product of the distributions of individual blocks (i.e., assuming block
independence). In Section \ref{sec3} we extend the LD-block model to incorporate
disease associated SNPs and epistasis and describe Monte Carlo Markov
chain (MCMC) algorithms to make inference from the joint model for both
LD-blocks and disease associations. In Section \ref{sec4} we briefly review our
previously developed Bayes factor-based test statistic, which is used to
further evaluate the statistical significance of candidate epistasis
detected by BEAM2, and discuss extensions of BEAM and BEAM2 for general
classification problems. In Section \ref{sec5} we demonstrate the superior
performance of BEAM2 by simulation studies and real data applications.
In Section \ref{sec6} we report our results from applying BEAM2 to the T2D data
from WTCCC (\citeyear{m07}). We conclude the article with a short discussion in
Section \ref{sec7}. More implementation details can be found in Supplemental
Material [\citet{ZhaZhaLiu}].

\section{A Bayesian model for LD-block inference}\label{sec2}

The data of interest consist of genotypes at a total of $L$ SNP markers
(or $L$ covariates, each taking on 3 possible values) observed in
$N_{d}$ case and $N_{u}$ control individuals. Let
$D=(D_{1},\dots,D_{L})$ denote a $N_{d}\times L$ matrix of case
genotypes, and $U=(U_{1},\dots,U_{L})$ denote a $N_{u}\times L$
matrix of control genotypes, where $D_{i}$ and $U_{i}$ denote
vectors of genotypes observed at SNP $i$ across individuals.

\subsection{Bayesian LD-block model}\label{sec2.1}

Here we introduce a Bayesian LD-block partition model without
considering disease association. Hence, we treat cases ($D)$ and
controls ($U)$ as coming from the same population and use the combined
data ($D,U$) to describe the model. A \textit{diplotype} for an
LD-block of SNPs is defined as a particular genotype combination of all
the SNPs in that block. We seek to partition the $L$ markers into $B$
consecutive blocks, so that the number of observed diplotypes within
each block is small (strong correlation), and correlation between SNPs
in different blocks is weak. Block partitions can be quite ambiguous in
many genomic regions and can vary across samples in details. Diplotype
block structures obtained from available software are often based on
ad hoc criteria that neither result in a proper uncertainty
measure nor optimize the association mapping power.\looseness=1

The block variable $B$ in our model consists of $L$ binary indicators
corresponding to the $L$ SNPs in the data. An indicator is equal to 1 if
the corresponding SNP is the start position of a block, and 0 otherwise.
As a result, $B$ uniquely defines a partition of SNPs into consecutive
blocks. For a diplotype block [$a,b)$ consisting of SNPs $a,\dots,b-1$,
we let ($n_{h}, m_{h})$ denote the counts of a particular
\textit{diplotype} $h$ in cases and controls, respectively. There are
3$^{b-a}$ possible diplotypes in the block. We assume that the diplotype
of each individual follows independently from a multinomial distribution
with frequency parameters \{$p_{h}\}$, and \{$p_{h}\}$ follows a
Dirichlet prior distribution, Dir(\{$\alpha_{h}\})$, where
\{$\alpha_{h}\}$ denotes a hyper-parameter (i.e., pseudo-counts). More
precisely, letting \{$n_{h}+m_{h}\}$ denote the combined counts of
diplotype $h$ observed in cases and controls, we have
\[
\Pr( \{ n_{h} + m_{h}\} |\{ p_{h}\} ) = \prod_{h = 1}^{3^{b - a}}
p_{h}^{n_{h} + m_{h}}\quad\!\!\mbox{and}\quad\!\!\Pr( \{ p_{h}\} ) = \frac{\Gamma (\alpha_\bullet)}{\prod_{h = 1}^{3^{b - a}} \Gamma (\alpha _{h})} \prod_{h = 1}^{3^{b
- a}} p_{h}^{a_{h} - 1},
\]
where the subscript `$\bullet$' denotes the sum of values over all subscripts.
We can integrate out $ \{p_{h}\}$ by
\begin{eqnarray*}
&&\int\Pr( \{ n_{h} + m_{h}\} |\{ p_{h}\} )\Pr( \{ p_{h}\} ) d( \{
p_{h}\} ) \\
&&\qquad = \frac{\Gamma (\alpha_\bullet)}{\prod_{h = 1}^{d - 1} \Gamma
(\alpha _{h})} \int\prod_{h = 1}^{3^{b - a}} p_{h}^{n_{h} + m_{h} +
\alpha _{h} - 1} d( \{ p_{h}\} ).
\end{eqnarray*}
Noting that the integrant on the right-hand side is proportional to the
density function of Dir($n_{h}+m_{h}+\alpha_{h})$ up to its normalizing
constant, we obtain the marginal probability of the combined data of
block [$a,b)$ as
\begin{eqnarray}\label{eq1}
&&P\bigl(D_{[a,b)},U_{[a,b)}|[a,b)\mbox{ forms a block}\bigr) \nonumber\\[-8pt]\\[-8pt]
&&\qquad = \Biggl( \prod_{h =
1}^{3^{b - a}} \frac{\Gamma (n_{h} + m_{h} + \alpha _{h})}{\Gamma
(\alpha _{h})} \Biggr)\frac{\Gamma (\alpha _{ \bullet} )}{\Gamma (n_{ \bullet}
+ m_{ \bullet} + \alpha _{ \bullet} )}.\nonumber
\end{eqnarray}
We set $\alpha_{h} = \rho /3^{b - a}$ for diplotype $h$, and by default,
we let $\rho = 1.5$. Based on the likelihood equivalence principle
[Heckerman, Geiger and Chickering (\citeyear{HecGeiChi95})], $\{\alpha_{h}\}$ is chosen
to be inversely proportional to the total number of possible diplotypes
in a block, and, hence, the sum $\alpha_{ \bullet}$ remains a constant
over different block sizes. Note that formula (\ref{eq1}) can also be used to
model~ca\-se data ($D)$ or control data ($U)$ only, which is simply done
by letting $m_{h}=0$, or $n_{h}=0$, respectively, for all $h$. Further
assuming independence between blocks (which is not entirely true, but
serves as a good approximation), the probability function $P(D,U|B)$ of
genotypes of all blocks can be expressed as the product of individual
block probabilities defined in formula (\ref{eq1}).

\subsection{Bayesian inference of SNP association based on
blocks}\label{sec2.2}

A saturated test of disease association for a diplotype block of $M$z
consecutive SNPs involves $3^{M} - 1$ free parameters. When $M$ is
moderately large ($M>3)$, the power of the test becomes exceedingly low.
We propose a Bayesian model of disease associations using only a subset
of markers in the block. Here, we only discuss joint associations for
SNPs within a diplotype block, and we will address epistatic
interactions in the next section. It therefore suffices to describe our
model for only one block.

Let $\{M\}$ denote the set of $M$ SNPs in the block. We assume that only
a subset $\{x\}$ of SNPs of size $x$ (often is 0 or 1) are truly
associated with the disease, and the SNPs in $\{ M\} \backslash\{ x\}$
are not associated with the disease given \{$x\}$. Thus, the diplotypes
of SNPs in \{$x\}$ are distributed differently and hence are modeled by
two different distributions for cases and controls, respectively.
Conditional on the diplotypes of SNPs in \{$x\}$, the diplotypes of SNPs
in \{$M\}\backslash\{x\}$, however, follow a common distribution between
cases and controls. The joint probability of the block data can
therefore be expressed~as
\begin{equation}\label{eq2}
P\bigl(D_{\{ M\}} ,U_{\{ M\}} \bigr)=P\bigl(D_{\{ x\}} \bigr)P\bigl(U_{\{ x\}} \bigr)P\bigl(D_{\{ M\}
\backslash \{ x\}} ,U_{\{ M\} \backslash \{ x\}} |D_{\{ x\}} ,U_{\{ x\}}\bigr).
\end{equation}
Here, $P(D_{\{ x\}} )$ and $P(U_{\{ x\}} )$ are modeled by two
independent multinomial-Dirichlet distributions specified in formula
(\ref{eq1}), treating all SNPs in \{$x\}$ jointly as a block. To model
$P(D_{\{M\}\backslash \{x\}},U_{\{M\}\backslash \{x\}}|
D_{\{x\}},U_{\{x\}})$, we combine the diplotypes of SNPs in $\{ M\}
\backslash\{ x\}$ in cases and controls together. These diplotypes are
not directly associated with the disease given \{$x\}$, and thus have
the same conditional distributions between cases and controls.
Conditional on each possible diplotype $h$ of SNPs in \{$x\}$, we model
the conditional diplotype distribution of SNPs in \{$M\}\backslash\{x\}$
again by a multinomial-Dirichlet distribution. It is straightforward to
derive the expression
\[
P\bigl(D_{\{ M\} \backslash \{ x\}} ,U_{\{ M\} \backslash \{ x\}} |D_{\{ x\}}
,U_{\{ x\}} \bigr) = P\bigl(D_{\{ M\}} ,U_{\{ M\}} \bigr)/P\bigl(D_{\{ x\}} ,U_{\{ x\}}
\bigr),
\]
where $P(D_{\{ M\}} ,U_{\{ M\}} )$ and $P(D_{\{ x\}} ,U_{\{ x\}} )$ are
specified in formula (\ref{eq1}), treating $\{M\}$ and $\{x\}$ as blocks,
respectively. We note, however, that the marker set $\{x\}$ is unknown
a priori, and needs to be inferred jointly with other
parameters in our Bayesian model.

\section{Joint inference of diplotype blocks and disease
association}\label{sec3}

\subsection{The joint model}\label{sec3.1}

To further incorporate epistatic interactions in formula (\ref{eq2}), and to
identify which SNPs are associated with the disease (response), we
partition all SNPs (not blocks) into three groups as in BEAM [Zhang and
Liu (\citeyear{ZhaLiu07})]. We introduce a latent $L$-dimensional indicator variable
($I)$ to represent the group memberships of the $L$ markers. For each
marker $i$, $I_{i}=0, 1, 2$ denotes three possible group memberships. SNPs
belonging to group-2 are assumed to be jointly associated with the
disease, that is, epistasis, which are modeled by two joint multinomial
distributions on the diplotypes over all group-2 SNPs---one for cases
and one controls. SNPs belonging to group-1 are assumed to be marginally
associated with the disease if they belong to different blocks, and are
modeled by mutually independent multinomial distributions conditional on
the case--control status and the block structure. If multiple group-1
SNPs fall into one block, we model their diplotypes jointly, that is,
group-1 SNPs within blocks become dependent of each other. If there are
both group-1 and group-2 SNPs within one block, we model the diplotypes
of group-1 SNPs within the block conditional on the diplotypes of the
group-2 SNPs within the block. SNPs belonging to group-0 are the
remaining SNPs unrelated to the disease status. We again model the
distribution of group-0 SNPs within a block by multinomial
distributions, with common parameters for cases and controls. We further
assume conditional independence of group-0 SNPs between blocks,
conditional on the group-1 and group-2 SNPs. More precisely, within each
block, we let $\{x_{2}\}$ denote the set of group-2 SNPs, let \{$x\}$
denote the union of the group-1 and group-2 SNPs, and let \{$M\}$ denote
all SNPs. We revise formula (\ref{eq2}) to take the form of a conditional
probability function:
\begin{eqnarray}\label{eq3}
&&P\bigl(D_{\{ M\} \backslash \{ x_{2}\}} ,U_{\{ M\} \backslash \{
x_{2}\}} |D_{\{ x_{2}\}} ,U_{\{ x_{2}\}} \bigr)\nonumber\\[-8pt]\\[-8pt]
&&\qquad =\frac{P(D_{\{ x\}} )P(U_{\{
x\}} )P(D_{\{ M\} \backslash \{ x\}} ,U_{\{ M\} \backslash \{ x\}}
|D_{\{ x\}} ,U_{\{ x\}} )}{P(D_{\{ x_{2}\}} )P(U_{\{ x_{2}\}} )}.\nonumber
\end{eqnarray}
Thus, group-0 and group-1 SNPs are no longer mutually independent as in
BEAM1, but are related to each other via the block structure. With
epistasis considered, the mutual independence between blocks in model
(\ref{eq2}) becomes conditional independence given group-2 SNPs. For notational
simplicity, we omit variable $I$ in (\ref{eq3}), but both $\{x\}$ and
$\{x_{2}\}$ are determined by $I$.

Given a particular block partition $B$ and SNP group memberships $I$, we
express the joint probability function of the entire case--control data
as
\begin{eqnarray}\label{eq4}
\qquad P(D,U|B,I) &=& P(D_{2}|B,I)P(U_{2}|B,I)\nonumber\\[-8pt]\\[-8pt]
&&{}\times\!\!\prod_{\{ M\} = [a,b) \in B}\!\!
P\bigl(D_{\{ M\} \backslash \{ x_{2}\}} ,U_{\{ M\} \backslash \{ x_{2}\}}
|D_{\{ x_{2}\}} ,U_{\{ x_{2}\}} ,B,I\bigr),\nonumber
\end{eqnarray}
where $D_{2}$ and $U_{2}$ denote the case and control genotypes of
group-2 SNPs, respectively, and the product term is defined in formula
(\ref{eq3}).

\subsection{Choice of prior distributions}\label{sec3.2}

We set the prior distribution of the block variable $B$ as the product
of independent Bernoulli probabilities $P(B)\,{=}\,\allowbreak p^{|B|}(1-p)^{L-|B|}$,
where $|B|$ denotes the sum of indicators in $B$. According to the block
distributions estimated in European and Asian populations by Gabriel et
al. (\citeyear{Gabetal02}), we assume that there are 50,000 blocks in the human genome
a priori, and thus we set $p = \min(0.5,50\mbox{,}000R/(3 \times
10^{9}L))$. Here,~$R$ denotes the length of the region spanned by the
$L$ SNPs, and $3\times10^{9}$ is the length of the human genome. A smaller
value of $p$ will help the method identify larger blocks, and a larger
$p$ will tend to identify smaller blocks. As the sample size (number of
individuals) increases, however, the impact of the prior choices
diminishes quickly. To avoid overfitting the blocks, we further impose a
restriction that the maximum number of observed distinct diplotypes in
a~block must be smaller than ($N_{d}+N_{u})/10$.

We set the prior distribution of the SNP membership variable $I$ as a
product of independent multinomial distributions, $P(I) = \prod_{i =
0}^{2} p_{i}^{| \{ j:I_{j} = i\} |}$, where $\{p_{0}, p_{1}, p_{2}\}$
denote the prior probability of each SNP belonging to group 0, 1, and~2,
respectively. By default, we set $p_{1} = p_{2} = \min(0.1, 5/L)$, and
$p_{0} = 1- p_{1} - p_{2}$. That is, we assume there are 10 SNPs
associated with the disease a priori, where~5 are marginally
associated with the disease, and 5 are associated through epistasis. Our
choice of the prior reflects that there are just a few SNPs truly
associated with the disease in a GWA study (where many other significant
SNPs are due to LD effects). Increasing this prior (and also increasing
the significance level) in the BEAM2 program may help identify
additional SNPs of moderate to low effects. To avoid overfitting in
interaction mapping, we further set an upper bound to the order of
interactions by $\ln_{3}((N_{d}+N_{u})/10)$. For example, when the sample
size is 1,000, our method can detect up to 4-way interactions. Overall,
changing the values of~$p_{1}, p_{2}$ may affect the posterior
distribution of SNPs in groups 1 and 2, but the effects will diminish as
the sample size increases.

Finally, the joint model of the observed genotype data in cases ($D$)
and controls ($U$), the block variable ($B$), and the SNP membership
($I$), is written~as
\begin{equation}\label{eq5}
P(D,U,B,I) = P(D,U|B,I)P(B)P(I),
\end{equation}
where the conditional distribution of ($D, U$) given ($B, I$) is
specified in formula~(\ref{eq4}).

\subsection{MCMC updates}\label{sec3.3}

The parameters of interest in our model are the block partition $B$ and
SNP membership $I$. We develop Metropolis--Hastings (MH) algorithms [Liu
(\citeyear{Liu01})] to update $B$, and, simultaneously, we develop a mix of a Gibbs
sampler and Metropolis--Hastings algorithm to update $I$. The posterior
distribution of ($B, I)$ is then output for further analysis.

To explore all possible block partitions, we propose the following
MH-mo\-ves: given a current block configuration $B$, we randomly select a
block~and
\begin{longlist}[(1)]
\item[(1)]divide the block into two new blocks at a random position;

\item[(2)]merge two adjacent blocks into one block; and

\item[(3)]randomly shift a block boundary to either left or right by $k$ SNPs,
where the shifting amount is constrained by other block boundaries.
\end{longlist}
The proposed move produces a new block partition $B'$, and the
move is accepted with probability $r = \min\{ 1,\frac{P(D,U,I,B')q(B' \to
B)}{P(D,U,I,B)q(B \to B')} \}$, where $q(B \to B')$ denotes the
probability of updating from $B$ to $B'$, and $P(D,U,B,I)$ is calculated
from the full model (\ref{eq5}). In our implementation, we chose the three types
of MH moves with probabilities 0.1, 0.1, and 0.8, respectively, and we
require a block to contain at least one SNP.

To update the SNP membership variable $I$, we updated the
membership~$I_{i}$ of SNP $i$ by calculating the posterior distribution of
$I_{i}=0,1,2$ given all other model parameters and the data. We also
propose a MH-move to switch the group memberships of two SNPs and accept
the move based on MH-ratios. Per MCMC iteration, we first run the Gibbs
sampler to update the memberships of all SNPs once, and then we run the
MH-sampler to switch each SNP in group-1 and group-2 once with SNPs in
other groups.

\section{Follow-up tests and generalization of the method}\label{sec4}

\subsection{A test of significance based on the Bayes factor}\label{sec4.1}

Although inference can be directly made from the posterior probabilities
output by BEAM2, the users may want to further evaluate the statistical
significance of the results in a frequentist way. In BEAM [Zhang and Liu
(\citeyear{ZhaLiu07})], we developed a novel Bayes factor, called B-stat, to evaluate
whether a SNP or a set of SNPs are significantly associated with the
disease, where the SNP set is selected by BEAM2 in our case.

For a set of $M$ SNPs to be tested, the null hypothesis is that all $M$
SNPs are not associated with the disease. Here, $M=1,2,3,\dots$
represents single-SNP, 2-way, and 3-way interactions, etc. B-stat for
the set of $M$ SNPs is defined as
\begin{equation}\label{eq6}
B_{M} = \ln\frac{P_{A}(D_{M},U_{M})}{P_{0}(D_{M},U_{M})} =
\ln\frac{P(D_{M})[ P(U_{M}) + \prod_{j \in M} P(U_{j}) ]}{P(D_{M},U_{M})
+ \prod_{j \in M} P(D_{j},U_{j})}.
\end{equation}
Here, $P_{0}(D_{M},U_{M})$ denotes the null genotype distribution (i.e.,
no disease association) at the $M$ SNPs in cases and controls, and
$P_{A}(D_{M},U_{M})$ denotes the alternative genotype distribution
(disease association). Under the null model, we assume that the
genotypes in both cases and controls follow the same distribution,
whereas under the alternative model, they follow different
distributions. We choose both $P_{0}(D_{M},U_{M})$ and
$P_{A}(D_{M},U_{M})$ as an equal mixture of two distributions: one that
assumes independence among the $M$ SNPs in controls (and also in cases
under the null model), which yields the product terms in formula (\ref{eq6}),
and the other that assumes a saturated joint distribution of all the $M$
SNPs. Note that the form of each term in~formu\-la~(\ref{eq6}) is defined in
formula (\ref{eq1}).

An interesting feature of B-stat is that it uses a mixture model to
accommodate the possibility that the $M$ SNPs may or may not be in
linkage equilibrium (independence). As a result, using B-stat will be
more powerful than using a standard likelihood ratio test or a
chi-square test of associations when the $M$ SNPs under testing are in
LD in controls.

We have previously shown that, under the null hypothesis of no disease
association, B-stat follows asymptotically a shifted chi-square
distribution with $3^{k}-1$ degrees of freedom [Zhang and Liu (\citeyear{ZhaLiu07})].
The shifting parameter can be computed explicitly, which is determined
by the sample size ($N_{d}, N_{u})$, the interaction size $M$, and the
Dirichlet hyper-parameter $\{\alpha_{h}\}$. Briefly speaking, the
shifting parameter is proportional to
$-(3^{M}-1)\ln(N_{d}N_{u}/(N_{d}+ N_{u}))$, and, thus, the larger the
number of individuals collected, or the more SNPs involved in an
interaction, the smaller the shifting parameter will be. In addition, if
large hyper-parameters $\{\alpha_{h}\}$ for the diplotype frequency
parameters are used, the shifting parameter will be large too. Note that
we want the B-stat to be small (e.g., $<$0) when the $M$ SNPs are not
associated with the disease, and, hence, the users should use small
values for $\{\alpha_{h}\}$, such as the default values we used in our
model.

\subsection{Generalization to classification problems with discrete
covariates}\label{sec4.2}

Let $\mathbf{Y}$ be the $n\times 1$ binary response vector, and let
$\mathbf{X}=(\mathbf{X}_{1},\dots,\mathbf{X}_{p})$ be the $n\times
p$ covariates matrix, with each covariate $\mathbf{X}_{j}$ taking on
$k_{j}$ discrete (ordinal or categorical) values. The standard
case--control genetic study setting can be viewed as using response
variables $\mathbf{Y}$ (i.e., case--control status) to fish out relevant
predictors $\mathbf{X}_{j}$ (i.e., SNPs). The epistasis mapping methods
BEAM attempt to find those $\mathbf{X}$'s that interactively affect
$\mathbf{Y}$. Both BEAM and the block-based method BEAM2 can be easily
extended to infer a classification model.

The idea of both BEAM [Zhang and Liu (\citeyear{ZhaLiu07})] and BEAM2 is to partition
the $p$ covariates in $\mathbf{X}$ into three nonoverlapping groups, such
that one group contains covariates unrelated with $\mathbf{Y}$, and the
other groups contain covariates either independently or jointly related
with $\mathbf{Y}$. The partition of the covariates is an unobserved latent
structure. Given a particular group partition of the covariates $I$, we
can compute $P(\mathbf{X}|\mathbf{Y}, I)$ as in Zhang and Liu (\citeyear{ZhaLiu07}),
which is analogous to that in a na\"{\i}ve Bayes model. BEAM2 further
segments the covariates $\mathbf{X}$ into ``blocks'' of highly correlated
variables, and treats blocks as mutually independent. This is achieved
by introducing a block indicator variable $B$, which is updated
iteratively together with the variable selection indicator $I$.

To predict the classification of a new observation with covariates
$\mathbf{x}_{\mathit{new}}$ based on the training data $(\mathbf{X}, \mathbf{Y})$,
we compute $P(\mathbf{X},\mathbf{x}_{\mathit{new}}|\mathbf{Y},\mathbf{y}_{\mathit{new}} =
1)$ and $P(\mathbf{X},\mathbf{x}_{\mathit{new}}|\mathbf{Y},\mathbf{y}_{\mathit{new}} = 0)$,
respectively, using the BEAM (or BEAM2) algorithm, and obtain the odds
ratio
\[
P(\mathbf{x}_{\mathit{new}}|\mathbf{X},\mathbf{Y},\mathbf{y}_{\mathit{new}} =
1)/P(\mathbf{x}_{\mathit{new}}|\mathbf{X},\mathbf{Y},\mathbf{y}_{\mathit{new}} = 0).
\]
The prior $P(\mathbf{y}_{\mathit{new}} = 1)$ can be estimated from the prior
knowledge of class distribution, such as the prevalence of a particular
disease in the population, which then leads to the posterior predictive
probability for $\mathbf{y}_{\mathit{new}} = 1$. A~computationally more attractive
way to do the computation is to output the latent variable partition and
block partition structures ($I$ and $B$ in our case) from their joint
posterior distribution inferred by the MCMC procedure of BEAM, and then
average the conditional odds over all the sampled~$I$ and $B$:
\[
P(\mathbf{x}_{\mathit{new}}|\mathbf{X},\mathbf{Y},I,B,\mathbf{y}_{\mathit{new}} =
1)/P(\mathbf{x}_{\mathit{new}}|\mathbf{X},\mathbf{Y},I,B,\mathbf{y}_{\mathit{new}} = 0).
\]
We tested this latter approach in a preliminary study and found the
results quite satisfactory.

The effect of BEAM2 is somewhat analogous to that of \textit{elastic
net} [Zou and Hastie (\citeyear{ZouHas05})] and \textit{group lasso} [Yuan and Lin
(\citeyear{YuaLin06})]. All methods attempt to address the phenomena that groups of
covariates tend to demonstrate associations with the response together,
and within groups the covariates are highly correlated. Different from
\textit{elastic net} and \textit{group lasso}, BEAM2 infers the
covariate groups and also the informative covariates within groups
jointly in a coherent probability framework. As a consequence, BEAM2
allows sparse variable selection at both the group level and the
individual variable level within groups, whereas \textit{elastic net}
and \textit{group lasso} do sparse selection only at the group level. In
a recent technical report, Friedman, Hastie and
  Tibshirani (\citeyear{FriHasTib}) attempt to achieve a
similar sparse selection effect as BEAM2 (sparse selection at both group
and individual levels) by introducing an additional penalty term.

Other important distinctions between BEAM2 (or BEAM) and those
lasso-based methods are the following: (a) the use of the na\"{\i}ve
Bayes framework to model $\mathbf{X}$ given $Y$ to greatly alleviate the
overfitting problem; (b)~the ability to incorporate interaction terms
without incurring a huge computational burden (with MCMC iterations);
and (c) the adoption of the Bayesian variable selection principle, which
is equivalent to using a more desirable $L_{0}$ penalty. The cost of
these advantages is that both BEAM and BEAM2 have to compute via MCMC
without a guarantee of always finding the optimal solution. Empirically,
however, the computational speed of BEAM and BEAM2 is no worse than that
required by lasso-type algorithms when the number of covariates is
large.

\section{Simulation studies and algorithm comparisons}\label{sec5}

\subsection{Block partition of HLA data}\label{sec5.1}

We first used the HLA region on human chromosome 6 to evaluate the block
partitions inferred by our method. The HLA region is one of the few
regions in the human genome in which recombination hotspots have been
experimentally verified [Jeffreys, Ritchie and Neumann (\citeyear{JefRitNeu00});
Jeffreys, Kauppi and Neumann (\citeyear{JefKauNeu01})]. We downloaded the genotype data
of 50 unrelated UK Caucasian semen donors from Jeffreys AJ's website.
The data covers a 216 kb region with 296 genotyped biallelic markers
spanning from the upstream of gene \textit{HLA-DNA} to gene
\textit{TAP2} in the MHC Class II region. It is known that this region
contains several prominent recombination hotspots [Jeffreys, Ritchie and Neumann
(\citeyear{JefRitNeu00}); Jeffreys, Kauppi and Neumann (\citeyear{JefKauNeu01})]. We therefore examined the relationship
between the experimentally verified recombination hotspots and the
SNP-block boundaries inferred by BEAM2. We used both haplotypes and
genotypes to evaluate our method, where the HLA haplotypes were first
inferred by CHB [Zhang, Niu and Liu (\citeyear{ZhaNiuLiu06})] from the genotype data. We
also simulated 1,000 individuals from the inferred HLA haplotypes using
HAPGEN [Marchini et al. (\citeyear{Maretal07})] to evaluate the performance of BEAM2
with a larger sample size. As a comparison, we applied HapBlock [Zhang et al. (\citeyear{Zhaetal02N2})], a dynamic programming-based algorithm of
block partitioning, to the same sets of data.

\begin{figure}

\includegraphics{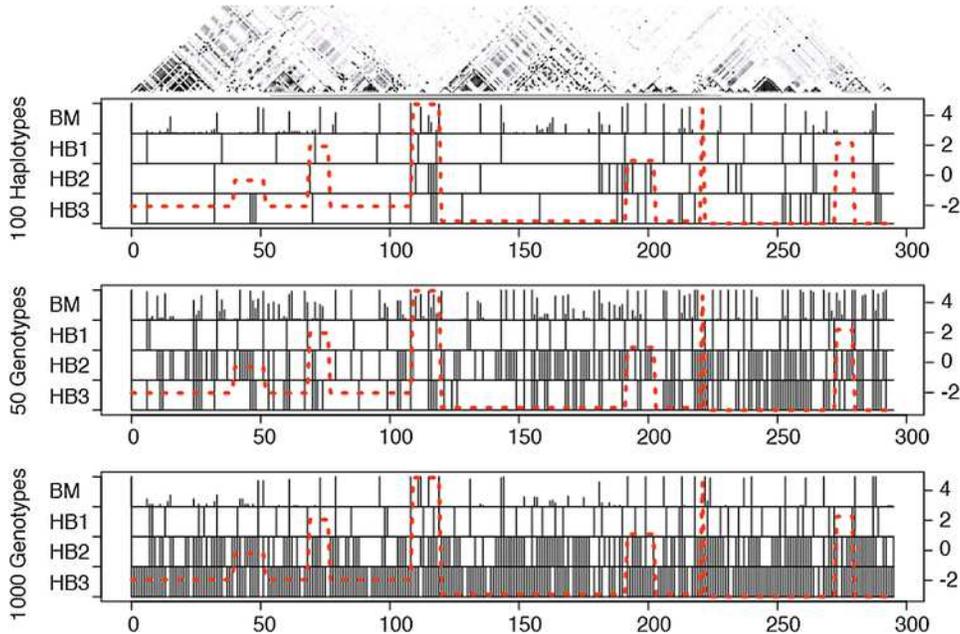}

  \caption{Block partition results on the HLA data. The top panel shows
the pairwise SNP LD calculated from 100 HLA haplotypes by Haploview. The
second, third, and forth panels show the block partition results by
BEAM2 (BM) and HapBlock (HB1, HB2, HB3) on three HLA data sets,
respectively. HB1 uses rule 1: common haplotypes; HB2 uses rule 2:
pairwise $D'$; and HB3 uses rule 3: four-gamete test. Experimentally
verified recombination rates are shown as dashed red lines (the unit
cM/Mb in the natural logarithm scale is shown on the right).}\label{fig1}
\end{figure}

With 100 haplotypes inferred from the 50 individuals, BEAM2 produced
accurate block partitions that correspond well to the visual blocks
displayed by Haploview [Barrett et al. (\citeyear{Baretal05})]. The block
boundaries also coincide with the known recombination hotspots within
the HLA region (Figure \ref{fig1}). It is further observed that, for the
haplotype data, the blocks inferred by BEAM2 are very similar to those
obtained by HapBlock. Unlike our model-based method, HapBlock requires
the user to specify ad hoc block partition rules, which can
result in undesirable partitions. We used three different rules to
define blocks: (1) common haplotypes, defined as a haplotype $>$5\% in the
sample, cover 80\% of samples in a block; (2) at least 80\% SNP pairs
with $D'>0.5$ in a block; or (3) four-gamete test on common haplotypes
($>$5\%) in a block. The blocks partitioned by HapBlock using each rule
are also shown in Figure \ref{fig1}.

Using the genotype data of the 50 individuals, we obtained very
different results between BEAM2 and HapBlock, and between the three
different rules of block partitions. Except for the first rule of
HapBlock, all other methods produced a large number of small blocks.
Small blocks generated by BEAM2 are due to the small sample size of 50
individuals, based on which the correlation between SNPs is hard to
detect using our likelihood model. The posterior probabilities of block
boundaries output by BEAM2, however, can be used as a measure of
uncertainty in block partitions. In comparison, HapBlock and many other
block partitioning methods only provide a single partition solution
without measuring block uncertainty.

Using the simulated genotypes of 1,000 individuals, we again observed
very different results shown in Figure \ref{fig1}. BEAM2 produced the cleanest
block partitions that corresponded well to the visual block boundaries
and to the known recombination hotspots. The $D'$ rule and the four-gamete
test rule via HapBlock again failed to produce reasonable partitions, of
which most blocks were singletons.

\begin{table}
\caption{Number of blocks inferred in HLA data sets}
\label{table1}
\begin{tabular*}{\tablewidth}{@{\extracolsep{\fill}}lccccc@{}}
\hline
\textbf{Data} & \textbf{BEAM2} & \textbf{BEAM2-10p} & \textbf{HapBlock-1} & \textbf{HapBlock-2} & \textbf{HapBlock-3}\\
\hline
100 haplotypes & 32 & 39 & 20 & \hphantom{0}30 & \hphantom{0}34\\
50 genotypes & 81 & 87 & 46 & 147 & 106\\
1,000 genotypes & 35 & 36 & 56 & 184 & 268\\
\hline
\end{tabular*}
\legend{``BEAM2-10p'' denotes BEAM2 applying a 10 times larger prior probability
than the default prior on the block boundary variable. HapBlock-1, 2, 3
denotes HapBlock applying three different block partition rules.}
\end{table}

We further show in Table \ref{table1} the number of blocks inferred by each method
in the three data sets. We observed that BEAM2 performed well (by which
we roughly mean that the number of estimated blocks is small, as is true
in the HLA region) in both the haplotype data and the 1,000 individuals'
genotype data. Because modeling genotypes (diplotypes) requires a much
larger set of parameters than modeling haplotypes, BEAM2 is expected to
perform worse in the 50 individuals' genotype data. As the number of
individuals increased to 1,000, however, our model-based approach
produced very similar partitions as that obtained in the haplotype data.
In comparison, HapBlock only preformed reasonably well in the haplotype
data, but produced many small blocks and singletons in the other two
data sets for all three block partition rules applied. HapBlock
performed the worst in the 1,000 individuals' genotype data, indicating
that the ad hoc rules applied by HapBlock do not produce
consistent block partitions as sample size increases. We further show in
Table \ref{table1} additional results by BEAM2 using a~10 times larger prior on the
block boundary variable, that is, we expect 10 times more blocks
a priori. We observed that the estimated posterior number of
blocks did not increase much, particularly in the 1,000 individuals'
genotype data, indicating that BEAM2 is insensitive to the prior choice
of block boundary variables. We also ran multiple MCMC chains to ensure
proper convergence.

\subsection{Simulation study using HapMap data}\label{sec5.2}

To mimic real genetic data observed in human populations, we first
randomly select a region in the human genome that contains 1,000 Illumina
HapMap 300k tagSNPs. The region also contains about the same number of
additional SNPs from HapMap PhaseII tagged by these tagSNPs, which we
refer to as nontagging SNPs. Two nontagging SNPs in the region are
randomly selected as disease SNPs. Given a disease model, we set the
marginal effect size (log odds ratio minus 1) per disease SNP at 0.5 and
choose a disease minor allele frequency (MAF) per locus from $(0.05,
0.10, 0.20, 0.50)$. Given a marginal effect size and a choice of MAF, we
then calculate the diplotype frequencies over the disease SNPs in cases
and controls, respectively [this is similarly done as presented in Zhang
and Liu (\citeyear{ZhaLiu07})]. According to the case--control diplotype frequencies
over the disease SNPs, we randomly sample 1,000 cases and 1,000 controls
from a pool of individuals without replacement. The pool consists of
10,000 control individuals generated by HAPGEN [Marchini et al. (\citeyear{Maretal07})]
using HapMap European sample (parents only) at odds $\mathrm{ratio}=1$, that is, no
disease association. Our simulation procedure is more economical than a
direct approach that generates one individual at a time and determines
its disease status conditional on the disease genotypes and penetrance,
because the direct approach may generate many more controls before
obtaining enough cases. Finally, we remove all nontagging SNPs from the
data including the two disease SNPs (which are typically unobserved in a
GWA study), and obtain a case--control data set containing 1,000 Illumina
HapMap tagSNPs.

\begin{table}[b]
\tablewidth=10cm
\caption{Disease models used in simulation study}
\label{table2}
\begin{tabular*}{\tablewidth}{@{\extracolsep{\fill}}lccc@{}}
\hline
\textbf{Risk} & $\bolds{A/A}$ & $\bolds{A/a}$ & $\bolds{a/a}$\\
\hline
Model 1\\
\quad $B/B$ & $1$ & $1+\theta$ & $(1+\theta)^{2}$\\
\quad $B/b$ & $1+\theta$ & $(1+\theta)^{2}$ & $(1+\theta)^{3}$\\
\quad $b/b$ & $(1+\theta)^{2}$ & $(1+\theta)^{3}$ & $(1+\theta)^{4}$\\[3pt]
Model 2\\
\quad $B/B$ & $1$ & $1$ & $1$\\
\quad $B/b$ & $1$ & $(1+\theta)^{2}$ & $(1+\theta)^{3}$\\
\quad $b/b$ & $1$ & $(1+\theta)^{3}$ & $(1+\theta)^{4}$\\[3pt]
Model 3\\
\quad $B/B$ & $1$ & $1$ & $1$\\
\quad $B/b$ & $1$ & $1+\theta$ & $1+\theta$\\
\quad $b/b$ & $1$ & $1+\theta$ & $1+\theta$\\
\hline
\end{tabular*}
\legend{Each table
cell lists the relative risk of the corresponding genotype combination.
Genotypes with risks equal to 1 have no effects to the disease. The
parameter $\theta$ is computed according to the specified marginal
effects (0.5 in our simulation) and disease MAFs $(0.05, 0.1, 0.2, 0.5)$.}
\end{table}

To evaluate the association mapping performance of our method, we
simulated case--control data sets based on the HapMap sample under three
disease models shown in Table \ref{table2}. Each disease model assumes 2 loci in
the genome contributing to the disease risk. While the first model
assumes no interactions, the other two models assume different types of
interactions. Using the simulated data sets, we compared the performance
of BEAM2 to BEAM1. We also implemented a third method that maps
associations and interactions based on predetermined block structures.
This third method serves as an intermediate method between BEAM1, which
is not block-based, and BEAM2, which infers block structures and maps
associations simultaneously. We used three different levels of
parameters (from stringent to liberal) to define blocks using existing
software, and we treated the diplotypes within each inferred block as
genetic alleles. The third method using three different block
definitions are hereafter referred to as Block1, Block2, and Block3,
respectively [more details of the third method can be found in the
Supplementary Material, \citet{ZhaZhaLiu}]. To compare the performance
of all methods, we ranked SNPs according to the association posterior
probabilities output by each method estimated for each data set. We then
calculated how often a method ranked the disease related SNPs among the
top SNPs. A SNP is regarded as being correctly identified as disease
related if it is within 5 SNPs on either side of a true disease locus.

\begin{figure}

\includegraphics{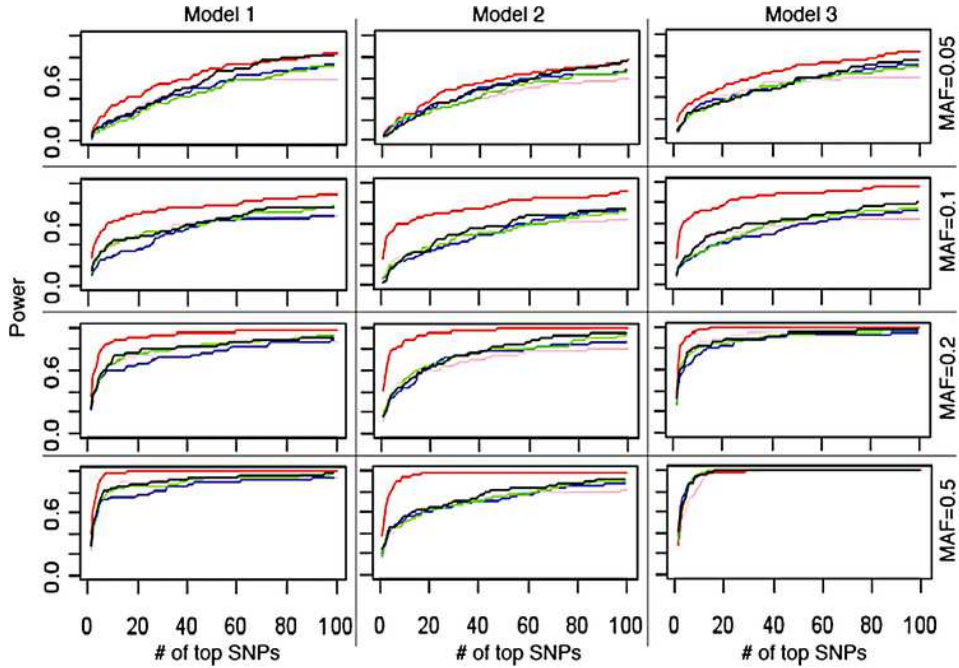}

  \caption{Power comparison for BEAM1, BEAM2, Block1, Block2, and Block3
using simulated data from the HapMap data. Under each simulation setting
and from 50 data sets, power (y-axis) is calculated as the proportion of
disease-associated SNPs (within 5 SNPs of true disease loci) among top
$m$ SNPs (x-axis), ranked by the posterior probability of association.
Each data set contains 1,000 candidate SNPs in 1,000 cases and 1,000
controls. The disease allele frequency is 0.05, 0.10, 0.20, and 0.50,
respectively. The marginal effect size of each disease SNP (unobserved)
is 0.5. Pink: BEAM1; Red: BEAM2; Blue: Block1; Green: Block2; Black:
Block3.}\label{fig2}
\vspace*{-3pt}
\end{figure}

As shown in Figure \ref{fig2}, under all parameter settings, BEAM2 performed the
best among all tested methods, where Block1, Block2, Block3, and BEAM1
all performed similarly. When disease allele frequency was low ($f=0.05$),
the power curves of all methods looked similar, but a closer inspection
of the top 5 ranked SNPs showed that BEAM1 only had $\sim $50\% chance
to capture disease related SNPs relative to BEAM2. When disease alleles
were common in the population ($f=0.10, 0.20, 0.50$), the advantage of the
BEAM2 model becomes obvious. Comparing the power curves for Model 2 and
Model 3, we observed that the power of BEAM2 increased much faster than
that of BEAM1 among the top 2 or 3 SNPs. We did not observe this
behavior in Model 1, which has no interactions. It thus indicates that
using SNP-blocks can increase the power of mapping both single SNP and
multi-SNP interaction associations. All methods compared here are
Bayesian methods that output posterior probabilities of disease
associations. It therefore indicates that our treatment of LD in BEAM2
is more appropriate than using either predetermined blocks or a Markov
chain model (as in BEAM1).

To further declare statistical significance, existing significance
estimation methods adjusting for multiple comparisons should be used,
such as the Bonferroni correction applied to B-stat introduced in BEAM1
[Zhang and Liu (\citeyear{ZhaLiu07})]. We compared BEAM2 with single SNP chi-square
tests using the above disease models [Supplementary Table S1, \citet{ZhaZhaLiu}], and observed that BEAM2 performed better than the
chi-square test for interaction Model 2 and Model 3, but performed the
same for the noninteractive Model 1.

\begin{figure}
\begin{tabular}{@{}c@{}}

\includegraphics{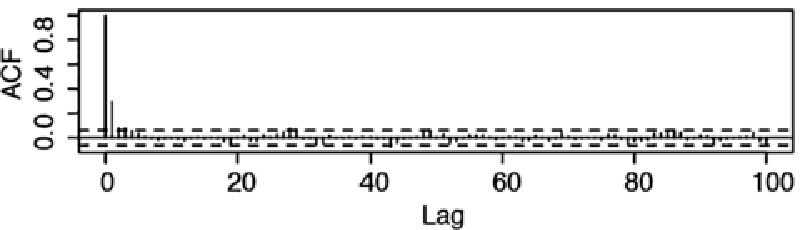}
\\
  (a)\\

\includegraphics{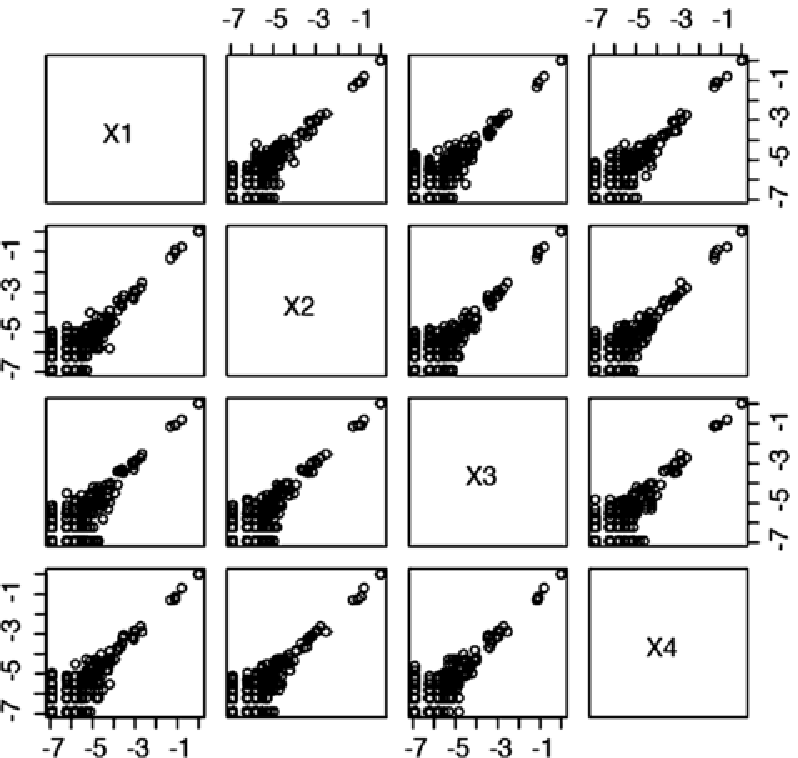}
\\
(b)
\end{tabular}
  \caption{Performance of MCMC. \textup{(a)} Autocorrelation plot obtained by
running BEAM2 on a simulated data set of 1,000 cases, 1,000 controls, and
1,000 SNPs under disease Mo\-del~2. A total of 1,000 iterations after
burn-in are used to calculate the plot. \textup{(b)} Posterior probabilities (in
logarithm scale) of disease association (marginal and epistatic) per SNP
compared across 4 independent runs of BEAM2.}\label{fig3}
\end{figure}

We also checked the performance of our MCMC sampling algorithm. As shown
in Figure \ref{fig3}, using a simulated data set from disease Model 2, the lag of
autocorrelation of our Markov chain is short, indicating fast
convergence of the Markov chain. We further compared the posterior
distribution of SNP associations from 4 independent runs of BEAM2, and
we observed close agreement between runs. In practice, the Markov chain
could converge to local modes, particularly if the data contain many
SNPs with complicated block structures. If block structures are of
primary interest, we suggest running BEAM2 in several runs to check if
the block partition results obtained in different runs are consistent.
More advanced MCMC algorithms, such as parallel tempering [Liu (\citeyear{Liu01})],
could further alleviate the local mode problems in MCMC sampling.

\begin{figure}
\begin{tabular}{@{}c@{\ \ }c@{}}

\includegraphics{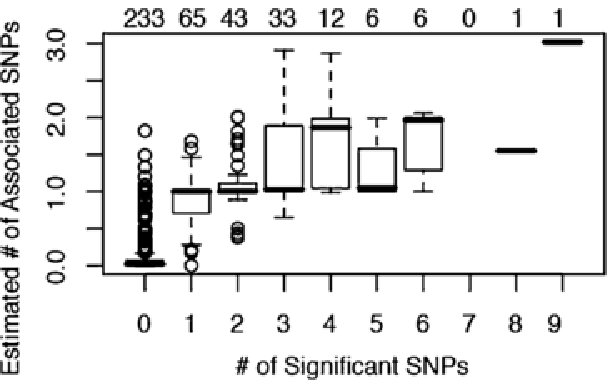}
&\includegraphics{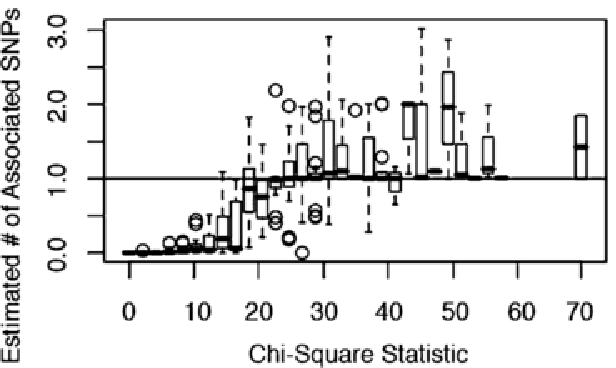}\\
(a)&(b)
\end{tabular}
  \caption{Estimated number of disease associated SNPs ($y$-axis) plotted
against \textup{(a)} the number of significant SNPs ($>$0.05 family-wise) and \textup{(b)}
the maximum single SNP chi-square statistics ($x$-axis) within a 100 kb
neighborhood per disease locus. In \textup{(a)}, the number of loci in each box
plot is further shown on the top. The plots are computed from 200
simulated data sets of disease Model 1.}\label{fig4}
\end{figure}

Using BEAM2, we can estimate the number of disease associated SNPs
around a disease locus by the sum of posterior probabilities of
associations over all SNPs within a neighborhood of a candidate locus.
Given our block-based association model, the number of associated SNPs
does not include SNPs whose disease association is merely created by LD,
and, hence, our estimates are more appropriate than a na\"{\i}ve count
of significant SNPs within the neighborhood. As shown in Figure \ref{fig4},
around a 100-kb neighborhood of every simulated disease locus in disease
Model 1, the estimated number of disease associated SNPs by BEAM2 is
around 1 when the association signal is sufficiently strong, even if
there are many significant SNPs in the neighborhood. The extra
significant SNPs created by LD make the localization of disease locus
difficult. This result highlights the importance of BEAM2 that performs
automatic variable selection within blocks. Rather than reporting
diluted small posterior probability of association over many neighboring
SNPs in LD, BEAM2 was able to select the strongest contributing SNP
within blocks (with large posterior probabilities of association) in our
simulation study.

\begin{figure}[b]

\includegraphics{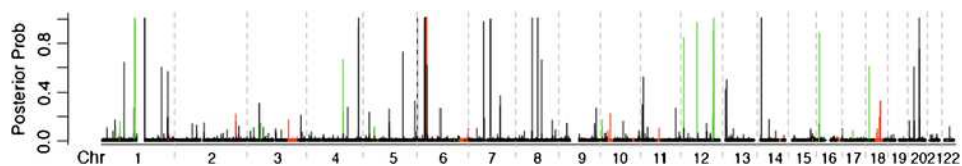}

  \caption{SNP-wise posterior probabilities ($y$-axis) of T1D associations
in 22 autosomal chromosomes ($x$-axis, chromosomes are separated by grey
dashed lines). Previously reported T1D associated genes are highlighted
in green, and candidate T1D associated genes, as defined at T1Dbase, are
highlighted in red.}\label{fig5}
\end{figure}

\section{Application to WTCCC type 1 diabetes data}\label{sec6}

We applied BEAM2 to analyze the T1D data set from the WTCCC project
[WTCCC (\citeyear{m07})]. The data set contains 2,000 T1D patients, 1,504 controls
from 1958 Birth Cohort (58C), and 1,500 additional controls from the
National Blood Service (NBS). Given our limited computation resources
(computation time, which would require several days to analyze half a
million SNPs in this data set; and memory usage, which would require
$>$4 Gb for half million SNPs), we applied BEAM2 to the top 10\% SNPs
ranked by marginal associations with T1D on all autosomes. We further
filtered out SNPs with bad clustering, SNPs violating Hardy--Weinberg
equilibrium in controls at 10$^{-5}$ significance, and ``almost
nonpolymorphic'' SNPs of which $>$95\% samples have the same genotypes.
The final data set contained 42,470 SNPs in 5,004 individuals. We ran
BEAM2 with 5 independent MCMC chains. Figure \ref{fig5} shows the averaged
posterior probabilities of T1D associations. Note that selecting top
10\% SNPs does not imply sparse and less correlated SNPs, because SNPs
in LD tend to be in or out of the top 10\% list together. Two
alternative ways to reduce the number of SNPs are to run BEAM2 on each
individual chromosome for chromosome-wise epistasis, or on previously
detected disease associated regions.

\subsection{Result summary and highlights}\label{se6.1}

We first compared our results with the SNPs reported in the original
WTCCC paper (\citeyear{m07}). As expected, we found that the SNPs reported by
BEAM2 and by the original WTCCC analysis are highly consistent. All
strongly associated SNPs reported in WTCCC are significant in our
analysis, and all strongly and moderately associated SNPs reported in
WTCCC have posterior probability${}>{}$0.1 by BEAM2 [Supplementary Table S2,
Zhang, Zhang and Liu (\citeyear{ZhaZhaLiu})]. We further compared our results with known T1D
associations obtained from T1Dbase (\href{http://www.t1dbase.org}{www.t1dbase.org}). Among the 55 SNPs
(or cluster of SNPs) output by BEAM2 with posterior probabilities
greater than 0.1, 17 (31\%) overlapped with known T1D associated
regions, including some well-known genes such as \textit{PTPN22} (1p13),
\textit{CTLA4} (2q33), MHC (6p21), and \textit{IL2RA} (10p15).

\begin{table}[b]
\caption{Strong single SNP associations with T1D}
\label{table3}
\begin{tabular*}{\tablewidth}{@{\extracolsep{\fill}}ld{5.14}ccc@{}}
\hline
\textbf{SNP} & \multicolumn{1}{c}{\textbf{Position}} & $\bolds{p}$\textbf{-value} & \textbf{T1Dbase} & \textbf{Gene}\\
\hline
rs6679677\tabnoteref{tab31} & \mbox{chr1:},\mbox{113.8--114.2 Mb} & 0 & Yes & \textit{PTPN22}\\
rs9405484 & \mbox{chr6:},\mbox{1.4 Mb} & 1.33e$-$8 & No & \textit{FOXC1}\\
MHC\tabnoteref{tab31} & \mbox{chr6:},\mbox{25--35 Mb} & 0 & Yes & \textit{MHC}\\
rs6592988 & \mbox{chr7:},\mbox{52.1 Mb} & 2.87e$-$7 & Yes & \textit{COBL}\\
rs11984645 & \mbox{chr8:},\mbox{55.2 Mb} & 7.06e$-$11 & No & \textit{MRPL15}\\
rs11782342 & \mbox{chr8:},\mbox{73.9 Mb} & 5.70e$-$12 & No & \textit{KCNB2}\\
rs11052552 & \mbox{chr12:},\mbox{9.7 Mb} & 2.61e$-$7 & Yes & \textit{CLEC2D}\\
rs11171739\tabnoteref{tab31} & \mbox{chr12:},\mbox{54.8 Mb} & 2.35e$-$11 & Yes & \textit{ERBB3}\\
rs17696736\tabnoteref{tab31} & \mbox{chr12:},\mbox{109.8--110.9 Mb} & 0 & Yes & \textit{CCDC63}, \textit{NAP1}\\
rs12924729\tabnoteref{tab31} & \mbox{chr16:},\mbox{11.1 Mb} & 1.01e$-$7 & Yes & \textit{CLEC16A}\\
\hline
\end{tabular*}
\legend{SNPs showing strong associations ($p\mbox{-value} < 5$e$-$7) with T1D by single SNP
test. $p$-value: nominal $p$-value of associations. T1Dbase: whether the
locus is documented in T1Dbase. Gene: nearest gene.}
\tabnotetext{tab31}{Additional SNPs in its neighborhood also show strong marginal
associations.}
\end{table}

In addition to the previously reported T1D genes, BEAM2 reported some
novel T1D associated loci. A list of likely T1D associations detected by
BEAM2, for both single SNPs (if $p\mbox{-value}<5$e$-$7) and two-way joint
associations (if $p\mbox{-value}<5$e$-$10), is shown in Tables \ref{table3} and \ref{table4},
respectively. For example, we detected 7 loci, among which two SNPs in
short distance form strong joint associations with T1D ($p\mbox{-value} <
5$e$-$10). These loci are not identifiable using single SNP tests, but
captured by BEAM2 as multi-SNP associations. We also found some likely
long-distance and cross-chromosomal interaction associations with T1D
($p\mbox{-value} < 5$e$-$11). One example is the joint association between SNP
rs3132676 in the classic MHC region on chromosome 6 and SNP rs9376523,
which is 111 Mb away on the same chromosome. This SNP pair is likely
interacting because their genotypes are strongly correlated in cases
(nominal $p$-value 8e$-$6 by test of independence), but not in controls
(nominal $p$-value 0.91). Although most two-way associations did not pass
the Bonferroni adjusted significance level in the genome scale, the
short-distance two-way associations are significant if only considering
local joint associations.

\begin{table}
\tabcolsep=0pt
\caption{Strong two-SNP associations with T1D}
\label{table4}
\begin{tabular*}{\tablewidth}{@{\extracolsep{\fill}}ld{5.3}cccd{5.3}ccc@{}}
\hline
\textbf{SNP1} & \multicolumn{1}{c}{\textbf{Pos1}} & \textbf{Pval1} & \textbf{Gene} & \textbf{SNP2} & \multicolumn{1}{c}{\textbf{Pos2}} & \textbf{Pval2} & \textbf{Gene} & \textbf{JointP}\\
\hline
rs7525703 & \mbox{chr1:},143 & 2e$-$2 & \textit{PRKAB2} & rs2077749 & \mbox{chr1:},143 & 3e$-$6 & \textit{PRKAB2} & 2e$-$12\\
rs6809441 & \mbox{chr3:},41 & 4e$-$2 & \textit{ULK4} & N/A & \mbox{chr3:},41 & 7e$-$2 & \textit{ULK4} & 5e$-$11\\
rs330483 & \mbox{chr4:},176 & 4e$-$6 & \textit{ADAM29} & rs330504 & \mbox{chr4:},176 & 1e$-$1 & \textit{ADAM29} & 2e$-$11\\
rs6906469 & \mbox{chr6:},10 & 1e$-$3 & \textit{DFCC1} & rs659964\tabnoteref{tab41} & \mbox{chr12:},111 & 2e$-$6 & \textit{ACAD10} & 4e$-$11\\
rs3132676\tabnoteref{tab41} & \mbox{chr6:},30 & 1e$-$5 & \textit{TRIM40} & rs9376523 & \mbox{chr6:},141 & 6e$-$4 & \textit{TXLNB} & 3e$-$11\\
rs9296661 & \mbox{chr6:},52 & 4e$-$4 & \textit{PKHD1} & rs1265566\tabnoteref{tab41} & \mbox{chr12:},110 & 1e$-$6 & \textit{CUTL2} & 5e$-$11\\
rs13340508 & \mbox{chr7:},75 & 9e$-$3 & \textit{CCL24} & rs17361077 & \mbox{chr7:},75 & 4e$-$5 & \textit{CCL24} & 6e$-$12\\
rs4838140 & \mbox{chr9:},124 & 2e$-$5 & \textit{NEK6} & rs7860360 & \mbox{chr9:},125 & 6e$-$4 & \textit{SCAI} & 4e$-$10\\
rs11104868 & \mbox{chr12:},87 & 1e$-$4 & \textit{KITLG} & rs7961663\tabnoteref{tab41} & \mbox{chr12:},110 & 3e$-$6 & \textit{CUTL2} & 1e$-$11\\
rs1958305 & \mbox{chr14:},23 & 5e$-$5 & \textit{DHRS2} & rs12100601 & \mbox{chr14:},23 & 2e$-$2 & \textit{DHRS2} & 0\\
rs7262414 & \mbox{chr20:},40 & 2e$-$5 & \textit{PTPRT} & rs2867064 & \mbox{chr20:},40 & 7e$-$2 & \textit{PTPRT} & 0\\
\hline
\end{tabular*}
\legend{Pairs of SNPs showing strong joint T1D associations ($p\mbox{-value} < 5$e$-$10) by
B-stat. If multiple SNP pairs are located around the same loci, only one
pair is shown. Pval1, Pval2, JointP represent nominal $p$-values of SNP1,
SNP2, and their joint associations, respectively.}
\tabnotetext{tab41}{SNP lies in known regions in T1Dbase.}
\end{table}

We further examined possible confounding effects of population
structures in the T1D data using a logistic regression model. The
regional information of WTCCC individuals is included as dummy
covariates. We observed that the test statistics of the detected SNP
associations remained almost unchanged before and after the adjustment
of population origins. We further randomly selected 10,000 SNPs
genome-wide and compared the distribution of their association
statistics with a chi-square distribution. The two distributions agreed
well [see Supplementary Figure S1, \citet{ZhaZhaLiu}]. We therefore
believe that population structure does not incur false positive
associations in the WTCCC T1D data.

\begin{figure}

\includegraphics{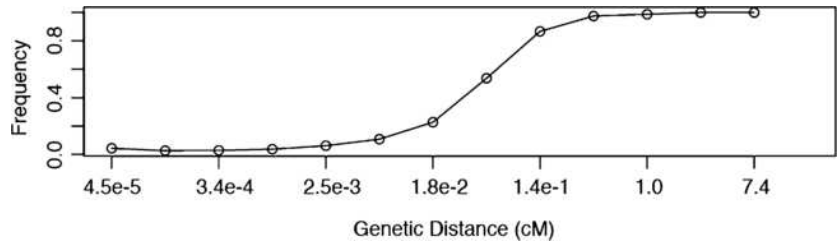}

  \caption{Frequency (y-axis) of block boundaries between adjacent SNP
pairs of certain genetic distance (x-axis), calculated from the T1D
data. The genetic distance is obtained from the HapMap CEU sample.}\label{fig6}
\end{figure}

We finally checked the block partition results of BEAM2 on the T1D data.
Given the large number of SNPs, we cannot visually inspect the blocks as
we did for the HLA data. Alternatively, we computed the genetic distance
between adjacent SNP pairs in the T1D data, using a genetic map\vadjust{\eject}
constructed from the HapMap CEU sample. Then, we checked how frequent
a~block boundary is inferred between SNP pairs in certain genetic
distance. Intuitively, the more distant two SNPs are located, the more
likely a block boundary occurs. As shown in Figure \ref{fig6}, our method
inferred almost 100\% block boundaries between SNP pairs with genetic
distance $>$1 cM, and the frequency of block boundary decreases as the SNP
pairs get closer. The block partitions between the 5 runs of BEAM2 are
consistent, with an average correlation coefficient of 0.96.

\subsection{Joint association patterns in the MHC region}\label{sec6.2}

T1D is an autoimmune disease and genes in the MHC region play an
important role in the immune system and autoimmunity
[Nejentsev et al. (\citeyear{Nejetal07}); Steenkiste et al. (\citeyear{Steetal})]. BEAM2
found a large number of SNPs within the MHC region showing extremely
strong association signals with T1D. Several multi-locus joint
associations within the MHC region are also detected. We therefore
examined more closely a 10-Mb MHC region, including the extended MHC
region (25--32 Mb) and the classic MHC region (32--35 Mb).

Within this 10 Mb region, we observed that the SNP pairs associated with
T1D are more often strongly correlated in cases than in controls [see
Supplementary Figure S2, \citet{ZhaZhaLiu}]. The joint associations
spanned from the classic MHC region to the extended MHC region over a
distance as long as 6.5 Mb. It has been previously reported that
haplotype blocks containing the most susceptible alleles
\textit{HLA-DRB1*03} and \textit{HLA-DRB1*04} within the
\textit{HLA-DR-DQ} region may extend as long as 2~Mb [Nejentsev et al.
(\citeyear{Nejetal07})] into the extended MHC region. It is thus arguable that the joint
associations observed between the MHC class II region and the extended
MHC region are due to extensive LD. We used a more traditional approach,
logistic regression, to test two-way interactions among SNPs within MHC
conditioning on the HLA-DR-DQ haplotypes. We first used CHB [\citet{ZhaNiuLiu06}] to infer haplotypes over \textit{HLA-DR-DQ} genes
(32.6--32.8 Mb). After collapsing rare haplotypes with frequencies lower
than 0.1\% into one group and obtaining 91 groups of haplotypes,\vadjust{\eject} we
regressed the T1D status on the 91 haplotype groups using 90 dummy
variables, which resulted in 78 significant haplotype groups over the
\textit{HLA-DR-DQ} genes explaining most of the MHC class II
associations with T1D (the 13 insignificant haplotype groups removed
from the model only changed the model deviance by 9.7).

\begin{figure}

\includegraphics{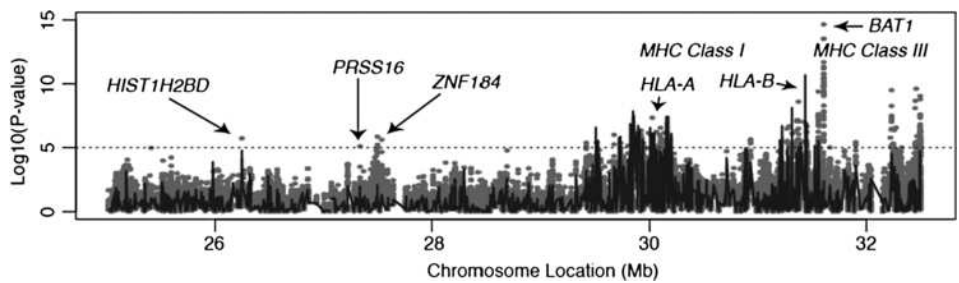}

  \caption{Log10 $p$-values ($y$-axis) of joint association test between SNPs
in the nonclass II MHC region ($x$-axis) and SNPs in the MHC class II
region (32.5--32.8 Mb), with main effects of the MHC class II SNPs
subtracted and conditioned on 78 T1D associated haplotypes in HLA-DR-DQ.
Each SNP has multiple $p$-values, shown as grey dots, corresponding to
interactions with different MHC class II SNPs. The black line for the
SNP indicates the $-\log10$ $p$-value of the nonclass II SNP's main effect.
A cutoff of 10$^{-5}$ for the $p$-values, shown as the dashed line,
corresponds to roughly one false positive among all tests.}\label{fig7}
\end{figure}

Conditioning on the 78 haplotype groups, we tested both main and
interaction effects between pairs of SNPs, one in the nonclass II region
(class I, class III, and the extended MHC region) and one in the class
II region. In particular, we regressed the T1D status on every pair of
SNPs and the 78 haplotype groups. The association statistic is the
change of deviance before and after including the two SNPs in the model.
We further subtracted the main effect of the class II SNPs. As shown in
Figure \ref{fig7}, we observed several peaks of $p$-values demonstrating
significant main and interaction effects within the nonclass II MHC
region. For example, we observed significant main effects of gene
\textit{HIST1H2BD} at 26.3 Mb (peak of black lines), interaction effects
of gene \textit{PRSS16} at 27.3 Mb and \textit{ZNF184} at 27.5~Mb (peaks
of grey dots), strong marginal effects of region 30--31.4 Mb including
class~I genes \textit{HLA-A} and \textit{HLA-B} (peaks of black lines),
and strong interaction effects of gene \textit{BAT1} at 31.6 Mb (peaks of
grey dots). The associations of \textit{PRSS16}, \textit{ZNF184}, and
\textit{BAT1} have been previously reported [Nejentsev et al. (\citeyear{Nejetal07});
Viken et al. (\citeyear{Viketal09})], and our analysis further
suggested that these genes are associated with T1D mainly through
interactive effects with MHC class II genes, controlling the MHC class
II haplotypes.

\section{Discussion}\label{sec7}

In this article we proposed a model-based Bayesian method for
simultaneous LD-block partitioning and multi-locus epistasis association
mapping. Different from many block-based methods, we combined block
partitioning and association mapping into a unified Bayesian model,
where both block structures and SNP associations are iteratively learned
through MCMC sampling. For block partitioning, our simulation study and
real data analysis showed that BEAM2 produces accurate and consistent
partitions of SNPs that compared favorably with other genetic knowledge
than some existing methods. The posterior probabilities of block
boundaries output by BEAM2 not only report the most likely block
partitions but also measure the uncertainty of blocks. Compared to some
existing methods using ad hoc partition rules, BEAM2 has two
main advantages: (1) it provides soft partitions rather than hard
partitions, that is, it provides a posterior distribution of block
partitions given the data, rather than a single block partition result;
soft partitions are necessary in regions with no dominant recombination
hotspots and block structures; and (2) it scales up well to large sample
sizes and produces consistent results. In particular, we showed that our
model-based method produces consistent block partitions as the number of
individuals increases, whereas the other methods we tested produced
drastically different results when more individuals from the same
population are included. For association mapping, BEAM2 is more powerful
than both the original BEAM algorithm, which accounts for LD using a
Markov chain, and the methods using pre-estimated blocks. BEAM2 tests
disease associations over uncertain block structures, where most SNPs
around disease loci are filtered out, as their associations are created
by LD with nearby disease SNPs.

The output of BEAM2 is a list of SNP-wise posterior probabilities of
marginal and interaction associations. From a frequentist point of view,
the identified SNPs can be further tested for genome-wide statistical
significance. We previously introduced a Bayes factor-based statistics,
B-stat [Zhang and Liu (\citeyear{ZhaLiu07})], for testing the significance of SNP
associations. B-stat performs similarly to a 2-df chi-square test for
single SNPs, but is more powerful for testing interactions of multiple
SNPs. The same B-stat can be used to test the statistical significance
of SNPs selected by the BEAM2 model, as we demonstrated in this paper.

When applied to the WTCCC T1D data, BEAM2 found all 5 statistically
significant loci previously reported in Table 3 of WTCCC (\citeyear{m07}), and
captured 7 moderately (insignificant) associated loci listed in Table 4
of WTCCC (\citeyear{m07}) with nontrivial posterior probabilities ($>$0.1). BEAM2
further reported some novel two-way joint associations, including 7 SNP
pairs ($p\mbox{-value}<5$e$-$10) in short-distance and 4 SNP pairs ($p\mbox{-value}<5$e$-$11)
in long-distance. The local two-way associations indicate main effects
of the related genes, which are, however, not detectable using single
SNP tests alone. The long-distance two-way associations did not pass the
genome-wide Bonferroni multiple testing control, but they may be
justifiable as real interaction associations if a better multiple
testing method is used and a~replication study is performed. We further
analyzed the well-known MHC region, and our analysis conditioning on the
MHC class II haplotypes suggested the existence of interaction
associations rather than MHC extended linkage alone. Given the complex
nature and the important biological role of MHC in the human genome, our
analysis is rather limited. Sophisticated analysis is in dire need to
further reveal the genetic mechanisms underlying MHC and immune
diseases.

The current BEAM model can be further improved in several ways. First,
missing genotypes and unobserved SNPs in the case--control sample can be
treated via imputation [Zhang (\citeyear{Zha})]. Previous studies have shown that
imputing untyped SNPs and missing genotypes from a reference panel can
improve the power of disease association mapping [Marchini et al. (\citeyear{Maretal07})]. Second, the current model only reports SNP-wise
posterior probabilities of associations to the disease without providing
a detailed analysis of association structures of the detected SNPs
interactions. We have observed in the MHC region a bulk of SNPs
exhibiting complex association structures with T1D. A post-analysis of
the MHC region is thus needed to delineate fine structures of the
selected SNPs and interactions with respect to their disease effects and
inter-relationships. Third, from a statistical point of view, it is
critical to control the false-discovery rate. This is particularly
important for multi-locus tests, which often involve a much larger
number of simultaneous comparisons than single SNP tests. We are
developing new statistical methods for evaluating genome-wide
statistical significance of associations.

\begin{supplement}
\stitle{Additional Supporting Information and Results\\}
\slink[doi,text={10.1214/11-AOAS469SUPP}]{10.1214/11-AOAS469SUPP}  
\slink[url]{http://lib.stat.cmu.edu/aoas/469/supplement.doc }
\sdatatype{.doc}
\sdescription{The
file includes a na\"{\i}ve SNP-block model used in our comparison,
verification of population structure in the sample, LD analysis of the
MHC region, Chi-square results of our simulation study, and comparison
of our results with previous results in the T1D WTCCC1 data.}
\end{supplement}

\section*{Acknowledgments}
We are grateful to three anonymous reviewers who helped us
substantially improve the manuscript from its original form.


\printaddresses

\end{document}